\documentclass[twocolumn,showpacs,preprintnumbers,amsmath,amssymb,floatfix,aps]{revtex4}
\begin{document}
\title{Charge and Lattice Dynamics of Ordered State in
    La$_{1/2}$Ca$_{1/2}$MnO$_3$: Infrared Reflection Spectroscopy Study}
\author{A.P. Litvinchuk$^1$, M. N. Iliev$^1$, M.~Pissas$^2$, and
C.W. Chu$^{1,3,4}$}
\affiliation{$^1$Texas Center for
Superconductivity and Advanced Materials, and Department of
Physics, University of Houston, Houston, Texas 77204-5002\\
$^2$Institute of Material Science, NCSR, Democritos, 153~10 Aghia
Paraskevi, Athens, Greece\\
$^3$Lawrence Berkeley National Laboratory, 1 Cyclotron Road,
Berkeley, California 94720\\
$^4$Hong Kong University of Science and Technology,
                        Hong Kong, China}
\date{June 6, 2004}
\begin{abstract}
We report an infrared reflection spectroscopy study of
La$_{1/2}$Ca$_{1/2}$MnO$_3$ over a broad frequency range and
temperature interval which covers the transitions from the high
temperature paramagnetic to ferromagnetic and, upon further
cooling, to antiferromagnetic phase. The structural phase
transition, accompanied by a ferromagnetic ordering at
T$_C$=234~K, leads to enrichment of the phonon spectrum. A charge
ordered antiferromagnetic insulating ground state develops below
the N\'eel transition temperature T$_N$=163~K. This is evidenced
by the formation of charge density waves and opening of a gap with
the magnitude of 2$\Delta_0$ = (320 $\pm$ 15) cm$^{-1}$ in the
excitation spectrum. Several of the infrared active phonons are
found to exhibit anomalous frequency softening. The experimental
data suggest coexistence of ferromagnetic and antiferromangetic
phases at low temperatures.
\end{abstract}
\maketitle
Manganite perovskites R$_{1-x}$A$_x$MnO$_3$ (where R is a
trivalent rare earth and A is a divalent alkaline rare earth)
exhibit rich phase diagram and a variety of intriguing properties
due the delicate interplay of spin, charge, lattice, and orbital
degrees of freedom (Ref.~\cite{review1,review2,piss03} and
references cited therein). Well defined anomalies of physical
properties at commensurate carrier concentrations of $x$=N/8
(where N=1, 3, 4, 5, and 7) and $x=L/3$ (where L=1,2) were
unambiguously established. Of particular interest is the
phenomenon of charge and orbital ordering, most clearly pronounced
for
$x$=1/2~\cite{chen96,rada97,mori98,huang,papa00,abra01,nale02}.

In this communication we report the results of an infrared
reflection study of La$_{1/2}$Ca$_{1/2}$MnO$_3$ over broad
frequency range and at temperatures, covering the transitions from
the high temperature paramagnetic to ferromagnetic and, upon
further cooling, to antiferromagnetic phase. By making use of
Kramers-Kronig analysis we obtained the spectral dependence of
conductivity. Its analysis yields information on the evolution of
phonons and electronic excitations as a function of temperature.
We found that additional phonon modes appear in the spectra in the
ferromagnetic phase, a fact which implies the occurrence of a
structural phase transition. Upon further cooling the conductivity
spectrum shows development of a gap, which signals the formation
of a charge ordered state at low temperatures. At the same time a
Drude-like component of the conductivity does not vanish
completely and suggests the coexistence of a metallic and
insulating phases.

The measurements were performed on a dense ceramic pellet of
La$_{1/2}$Ca$_{1/2}$MnO$_3$, mechanically polished to optical
quality. The material preparation technique is described in
Ref.~\cite{piss03,abra01,simo00}. The samples were intensively
characterized by the X-ray scattering, neutron diffraction,
magnetization measurements, Raman scattering as well as
M\"ossbauer spectroscopy (doped with 1\% Sn or Fe). These
measurements unambiguously identified the transition from the
paramagnetic to ferromagnetic phase at T$_C$=234~K, and further to
antiferromagnetic phase at T$^c_N$=163~K (cooling cycle) or
T$^h_N$=196~K (heating cycle), see Fig.~6 in Ref.\cite{piss03}.

The reflection measurements were performed on a Bomem-DA8
Fourier-transform interferometer in the frequency range 50-8000
cm$^{-1}$ with the use of a liquid-helium-cooled bolometer,
HgCdTe, and InSb detectors and appropriate beam splitters.
Spectral resolution was set to 1~cm$^{-1}$. A gold mirror was used
as a reference. The sample was attached to the cold finger of a
helium flow cryostat and during measurements the temperature was
stabilized to within 0.2~K. The reflectance spectra R($\omega$)
were extrapolated in the low frequency range by either
Hagen-Rubens relation (1-R)~$\sim$~$\omega ^{1/2}$ or a constant
(at low temperatures) and R~$\sim \omega^{-4}$ for high
frequencies up to 30000 cm$^{-1}$. The results of Kramers-Kronig
analysis of the spectra show that variation of extrapolated
reflection do not influence conductivity values in the frequency
range of interest (50-3000 cm$^{-1}$).

The reflectance spectra of La$_{1/2}$Ca$_{1/2}$MnO$_3$ during
cooling and heating cycles are shown in Fig.~1 for several
temperatures between 215 and 110~K. As expected, a hysteretic
temperature behavior is clearly seen around T$_N$, typical for the
first-order transition occurring in the CE-type magnetic
structure\cite{woll55}. Indeed, upon cooling the low-frequency
reflectance (below 200 cm$^{-1}$) shows a "metal-like" increase
toward lower wavenumbers for the four upper curves on the left
panel of Fig.~1, but becomes less frequency dependent for the two
lower curves (i.e. below 145~K). Instead, during the heating cycle
the low-frequency reflectance keeps its behavior from the low
temperatures up to 190~K and becomes more "metal-like" at higher
temperatures only.

To obtain more specific information on phonons and the charge
dynamics, we performed Kramers-Kronig analysis, which yields the
spectral dependence of conductivity. Fig.~2 illustrates the data,
obtained in the cooling cycle. The conductivity spectra below 650
cm$^{-1}$ are dominated by phonons, while at higher frequencies
the "background" conductivity steadily increases toward higher
wavenumbers with apparent slope becoming larger upon lowering
temperature. At T=80~K (the lowest panel in Fig.~2) the
extrapolation of this background to lower frequencies crosses zero
conductivity at positive wavenumbers, i.e. shows zero contribution
to the {\it dc} conductivity. It signals opening of a gap, related
to the formation of a charge density wave in charge ordered state
due to real-space ordering of Mn$^{3+}$ and Mn$^{4+}$
ions\cite{review1,chen96,calv98}.

Theoretical consideration of a charge ordered system yields the
following frequency dependence of conductivity\cite{lee74}:

\begin{equation}
\sigma(\omega) \sim (\omega - 2\Delta)^{\alpha},
\end{equation}

\noindent where $2\Delta$ is the magnitude of the charge gap and
$\alpha~=~1/2$. Using this prediction we fitted the conductivity
spectra at frequencies above the highest energy phonon (in the
range 750 to 2800 cm $^{-1}$) using $2\Delta$ and $\alpha$ as
parameters. For all temperatures the values of $\alpha$ are found
to be between 0.51 and 0.54, in good agreements with the theory.
The parameter $2\Delta$ increases in a linear manner from the
negative value of $-2777$ cm$^{-1}$ at room temperature to $-171$
cm$^{-1}$ at T=170~K and becomes positive at 160~K ($2\Delta$ =
165 cm$^{-1}$), signaling opening of a "real" gap. The
temperature, at which this gap opens, unambiguously identifies it
as being due to the formation of a charge ordered state in
La$_{1/2}$Ca$_{1/2}$MnO$_3$ because the antiferromagnetic
transition temperature for the sample is T$^c_N$=163~K. The gap
fully opens below 150~K, where it reaches the value of
$2\Delta_0$=(320~$\pm$~15) cm$^{-1}$ (Fig.~3).

Earlier experimental study of the charge density waves in
La$_{1/2}$Ca$_{1/2}$MnO$_3$ by optical transmission technique
yielded the gap value of 710 cm$^{-1}$~\cite{calv98}. Even larger
value of about 3600 cm$^{-1}$ was obtained from a reflection
studies by Kim et al.\cite{kim02}. Unlike present measurements,
which were performed on a bulk sample, the measurements in
Ref.~\cite{calv98} we carried out on pressed pellets of finely
milled La$_{1/2}$Ca$_{1/2}$MnO$_3$, embedded into CsI host matrix.
We believe that there are at least two factors, which contributed
to an overestimation of $2\Delta$ in this latter case. First,
pellets non-uniformity may introduce scattering of the transmitted
light beam and this way increase apparent optical density of the
sample. Second, the authors of Ref.~\cite{calv98} performed
fitting of the "background" in a very narrow frequency interval
710-900 cm$^{-1}$, which could generate significant error in
determining the slope and, correspondingly, the value of
$2\Delta$. As to the results of Ref.~\cite{kim02}, the
measurements were performed in a wide frequency range extending up
to 30~eV with emphasis on the analysis of an intense feature near
1~eV (due to an interatomic Mn$^{3+} \rightarrow$ Mn$^{4+}$
transitions\cite{jung98}); the authors overlooked evolution of the
spectra at lower wavenumbers and lower conductivity values.

Next, we turn to the analysis of phonons, which dominate
conductivity at frequencies below 650 cm$^{-1}$. The spectra for
several temperatures above and below charge ordering temperature
T$^c_N$=163~K are shown in Fig.~4. Each spectrum is fitted by a
set of Lorentzians, which correspond to phonons. The contribution
of free carries is accounted for by an additional oscillator,
centered at zero frequency. As it is seen, with just 5 phonons one
can adequately describe the spectrum at 230~K. At 215~K two new
bands, centered at about 505 and 285 cm$^{-1}$, appear. Intensity
of these bands gradually increases upon lowering temperature down
to 150~K,  and then becomes weakly dependent on temperature upon
further cooling (Fig. 5(a)). These new lines in the spectra could
be a consequence of either the formation of a novel orthorhombic
phase below T$_C$=234~K, which has the same symmetry (space group
{\it Pnma}), but slightly different lattice parameters compared to
the room temperature phase of
La$_{1/2}$Ca$_{1/2}$MnO$_3$\cite{huang}, or the appearance of a
superstructure with doubled $a$ lattice parameter and the space
group {\it P2$_1$/m}\cite{rada97}. The observed intensity increase
over rather wide temperature interval below T$_C$ indicates that
the volume fraction of the novel phase increases upon sample
cooling, in agreement with\cite{huang}. Note that appearance of
vibrational modes with very similar frequencies was reported in
charge ordered (LaPrCa)MnO$_3$~\cite{lee02}.

The number of phonon lines observed in the infrared spectra is
small compared to what is predicted by a group theoretical
analysis\cite{abra01,smir99}: 25 for the room temperature phase of
La$_{1/2}$Ca$_{1/2}$MnO$_3$ (space group {\it Pnma}) and 63 for
the low-temperature charge ordered phase (space group {\it
P2$_1$/m}). For the parent compound LaMnO$_3$, as shown in the
upper panel of Fig.~4, one clearly identifies majority of
theoretically predicted lines, as also reported by Paolone et
al.\cite{paol00} and Quijada et al.\cite{quij01}. The small number
of lines observed in La$_{1/2}$Ca$_{1/2}$MnO$_3$ is probably due
to the effect of compositional cation disorder (La/Ca), which
considerably shorten the phonon lifetime and, consequently,
broaden phonon peaks. Indeed, the typical phonon line width in
LaMnO$_3$ is 10-40 cm$^{-1}$ at room temperature, while it is as
high as 60-80 cm$^{-1}$ in La$_{1/2}$Ca$_{1/2}$MnO$_3$. The effect
of phonon line broadening is documented in a study of 8\% Ca-doped
LaMnO$_3$. \cite{paol00}

Another interesting experimental finding is that several phonon
lines exhibit pronounced frequency softening upon entering the
charge ordered insulating antiferromagnetic state (Fig.~5 (c,d)).
This could be due to the variation of relevant bond distances,
documented in the neutron diffraction studies\cite{rada97,huang}
and/or the effect of magnetic order on corresponding force
constants, similar to those reported for other magnetic
materials\cite{home95,litv04}.

It is important to note that the dielectric response of
La$_{1/2}$Ca$_{1/2}$MnO$_3$, as obtained from the reflection
spectroscopy data, is consistent with the phase separation
scenario for manganese oxides (see review \cite{dago01} and
references cites therein). Indeed, even below T$_N$ the dielectric
function is not typical of an insulator, but contains the
Drude-like component, a signature for presence in the sample of a
conducting phase. Fig.~5(b) shows the temperature dependence of
integrated conductivity in the frequency range 50-650 cm$^{-1}$
after high-frequency "background" was subtracted. It contains
contribution of phonons and low-frequency metallic response. As
phonons are only weakly change with temperature (see Fig.~4), this
dependence reflects primarily the fraction of conducting
(ferromagnetic) phase in the sample: it sharply increases upon
entering ferromagnetic state below T$_C$, reaches its maximum
around T$_N$, but does not completely disappear at lower
tempe\-ra\-tu\-res. This finding is in agreement with earlier
reports on La$_{1/2}$Ca$_{1/2}$MnO$_3$
\cite{simo00,smol00,roy00,frei02,tong03}. We have to mention that
the presence of a conducting phase has only minor effect on the
gap magnitude of the charge ordered state, as determined above.
This is due to the fact that the conducting phase affects the
dielectric function at low frequencies (below 300 cm$^{-1}$),
while the gap magnitude determination involves analysis of the
spectra at higher frequencies (above 750 cm$^{-1}$). The different
value of the gap compared to that of Ref.~\cite{calv98} could, at
least in part, be due to different size of charge ordered domains
in the material.

In conclusion, the infrared reflection study of
La$_{1/2}$Ca$_{1/2}$MnO$_3$ revealed the occurrence of a
structural phase transition at T$_C$, evidenced by appearance of
additional phonon lines. The charge and orbital ordering below the
antiferromagnetic transition temperature T$_N$ was found to
drastically modify the carrier dynamics as charge density waves
develop, leading to the creation of a gap in the excitation
spectrum with magnitude of 2$\Delta_0$=(320 $\pm$ 15) cm$^{-1}$.
The experimental data suggest coexistence of ferromagnetic and
antiferromangetic phases at low temperatures.

\acknowledgments The work at the University of Houston is
supported in part by the State of Texas through the
    Texas Center for Superconductivity and Advanced Materials, NSF
    grant No. DMR-9804325, the T.L.L.~Temple Foundation, the J.J. and
    R.~Moores Endowment. At LBNL we acknowledge the support by the Director,
                Office of Energy Research, Office of Basic Energy Sciences,
                Division of Materials Sciences of the US Department of Energy
                under contract No. DE-AC03-76SF00098.

\begin{figure*}
\caption{Reflection spectra of La$_{1/2}$Ca$_{1/2}$MnO$_3$ as a
function of temperature during cooling (left panel) and heating
(right panel) cycles. The vertical scale corresponds to the lowest
spectrum in each panel, and the other spectra are consequently
shifted by 0.1 for clarity. Arrows mark the position of phonon
modes, which appear in the spectra at low temperatures (see the
text for details). }
\end{figure*}
\begin{figure*}
\caption{Temperature dependent conductivity of
La$_{1/2}$Ca$_{1/2}$MnO$_3$, as obtained form the reflection in a
cooling cycle. Points are experimental data, dotted lines
represent Drude oscillator and high-frequency conductivity
component, extrapolated to lower energies (see text for details).
Solid curves, along with the two mentioned terms, include
contribution of optical phonons. Note that the extrapolation of
high frequency "background" crosses zero line in the two lower
panels, indicating the opening of a gap.}
\end{figure*}
\begin{figure*}
\caption{ The charge density waves gap magnitude $2\Delta$ as a
function of temperature. Line is a guide to the eye.}
\end{figure*}
\begin{figure*}
\caption{Experimental conductivity spectra of
La$_{1/2}$Ca$_{1/2}$MnO$_3$ (points) as a function of temperature
in a cooling cycle and their fit (solid lines) with Lorentzian
oscillators, representing phonons and Drude component (dashed
lines). The upper panel shows room temperature conductivity
spectrum of the parent LaMnO$_3$ compound, where a large number of
relatively narrow phonon lines is observed. }
\end{figure*}
\begin{figure*}
\caption{Temperature dependence of the integrated low-frequency
conductivity (b), line intensity (a) and peak position (c,d) for
several phonon modes of La$_{1/2}$Ca$_{1/2}$MnO$_3$. Solid lines
in (c) and (d) show the temperature dependent position expected
for a standard anharmonic phonon decay, while dotted lines in
(a,b) are guides to the eye.}
\end{figure*}
\end{document}